# On narrowing coated conductor film: emergence of granularity-induced field hysteresis of transport critical current

A. A. Gapud[a], D. K. Christen[b], R. Feenstra[b], F. A. List III[b], A. Khan[a],

[a]*University of South Alabama, 307 University Blvd, Mobile, AL 36688*
[b]*Oak Ridge National Laboratory, 1 Bethel Valley Rd, Oak Ridge, TN 37831-6061*

## Abstract

Critical current density $J_c$ in polycrystalline or granular superconducting material is known to be hysteretic with applied field $H$ due to the focusing of field within the boundary between adjacent grains. This is of concern in the so-called coated conductors wherein superconducting film is grown on a granular, but textured surface of a metal substrate. While previous work has mainly been on $J_c$ determined using *induced* or magnetization currents, the present work utilizes *transport* current via an applied potential in strip geometry. It is observed that the effect is not as pronounced using transport current, probably due to a large difference in criterion voltage between the two types of measurements. However, when the films are *narrowed* by patterning into 200-, 100-, or 80-μm, the hysteresis is clearly seen, because of the forcing of percolation across higher-angle grain boundaries. This effect is compared for films grown on ion-beam-assisted-deposited (IBAD) YSZ substrate and those grown on rolling-assisted-biaxially-textures substrates (RABiTS) which have grains that are about ten times larger. The hysteresis is more pronounced for the latter, which is more likely to have a weak grain boundary spanning the width of the microbridge. This is also of concern to applications in which coated conductors will be striated in order to reduce of AC losses.

1. Introduction

In the push to develop superconducting films of YBa$_2$Cu$_3$O$_x$ as the medium for second-generation power-transmission technology, practitioners have taken the practical path of depositing such film on long lengths of metallic tape. Since typical tapes are inevitably polycrystalline, great progress has been made in *texturing* this substrate so that grain-to-grain misalignment is minimized, down to as small as about 2 degrees – which is the maximum misalignment below which there is no detrimental effect on critical current [1, 2] An example of these so-called *biaxially textured* substrates are rolling-assisted biaxially textured substrates (RABiTS) and ion-beam-assisted-deposition yttria stabilized zirconia (IBAD YSZ) [3-6]. In both cases, oxide buffer layers ensure epitaxial growth of and chemical compatibility with YBCO, resulting in the so-called "coated conductor" architecture.

In the interest of reducing AC losses by subdividing the single wide strip of HTS coating into an array of narrow filaments [7], it is important to ascertain any effects on transport properties, especially the critical current density $J_c$ as a function of applied field $H$. Since in the overall orientation distribution there remain some grain misalignments large enough to limit *local* current levels, one could expect $J_c$ to be affected by tracks that are not significantly wider than the typical grain size [8, 9] . For currents limited by high-angle boundaries, however, there is the possibility that flux *trapped* within grains could partially cancel out the applied field within grain *boundaries,* producing a hysteretic effect on $J_c$ as a function of $H$. This effect has been long observed in polycrystalline superconductors [10], wherein $J_c(H)$ is monotonic for increasing $H$ and becomes non-monotonic for $H$ decreasing from above a threshold value and where maximum $J_c$ occurs at nonzero $H$.  This phenomenon is generally accepted to be a consequence of flux trapping within the grains, which when no longer supported by applied field, produces a 'focused field' at the grain boundaries (GB's) because of oppositely-directed intragrain currents on either side of the GB.  The minimum local field resulting from the partial cancellation between $H$ and the focused field effectively shifts the maximum of $J_c(H)$ to a nonzero $H$..  Recently, this has been illustrated with insightful and systematic studies in model systems with a *single* GB [11, 12].  The asymmetric

"peak" in $J_c$ for decreasing $H$ was clearly shown to occur only when magnetization currents were made to cross a grain boundary, and the value of $H$ at which the peak occurs was shown to be closely correlated with the *intra*granular $J_c$ at $H = 0$, thus strongly supporting a connection with intragranular flux trapping. Building on this, Palau *et al.* [13]showed the hysteresis effect as a possible, non-invasive method for analytically separating the intergranular and intragranular critical currents in YBCO coated conductors, wherein there is great interest in determining the extent to which current-carrying capacity is limited by the bulk material in the grains or by the linkages between the grains.

The approach of ref 13 has proven successful for features of the *magnetization $J_c$*, determined by the combination of large-scale percolating persistent currents and the circulating strong currents within each grain. However, in applications such as power transmission, the dominant current is "forced" by applied potential across the superconducting film – i.e., *transport* current. It is therefore of interest to probe granularity by this method, although there have been only limited reports of hysteresis in *transport $J_c(H)$* for coated conductors [14]. Although in both measurements the effects of percolation around high-angle grain boundaries (GB) should be identical [15-17], the hysteresis phenomenon is most evident in the magnetization measurements. We conjecture that the reason may stem from the fact that the electric field levels of the two approaches are typically different by several orders of magnitude (e.g., 1 µV/cm transport criterion vs. ~$10^{-11} – 10^{-13}$ V/cm for SQUID magnetometry)[18, 19]. Qualitatively, the higher voltage levels in transport drive some current across boundaries of intermediate strength . In the magnetic case, much less current is forced through these same moderate GBs, leading to effectively more isolated grains, which support field-history driven intragrain circulating currents. Since it is these intragrain currents that lead to the GB field focusing effects, resultant history-induced GB conduction is observed more prominently in the low-voltage magnetization measurements. A more quantitative assessment of this argument is beyond the scope and intent of this paper.

If a film were to be *narrowed,* it is possible to enhance the effect of high-angle grain boundaries by reducing or eliminating percolative path options [8, 9].. As a result, a larger fraction of sample comprises current-limiting boundaries and therefore grains that support only strong, circulating intragrain currents. It is these grains that promote the enhancement of intergrain currents when the GB field is nearly cancelled. In an extreme case, the effect could approach that seen for a *single*-GB[11, 12, 20-23]. The present work shows that hysteresis of transport $J_c(H)$ is indeed observed after a film is narrowed.

**2. Experiment**

All films in this study were fabricated using the $BaF_2$ *ex situ* method which is known for reproducibly high-quality films, as described in detail elsewhere[24]. At 77 K, $J_c$ for these films is typically ~ 3 MA/cm$^2$ (full-width/unpatterned) and irreversibility field is typically ~ 7 T. In order to probe the effect of grain size, one set of films were deposited on rolling-assisted biaxially textured substrates (RABiTS) – which are known to produce grain sizes of 10 to 50 μm -- and another on ion-beam-assisted-deposition yttria stabilized zirconia (IBAD YSZ) – which are known to produce smaller grain sizes on the order of a μm [25]. By photolithography and wet etching, a 1.1-μm-thick film on RABiTS and a 0.3-μm-thick film on IBAD YSZ each was patterned into 100-μm parallel strips spanning the length of the sample, separated by 100-μm gaps. Using laser scribing, a 200-μm x 1-mm bridge was patterned onto a 1.1-μm-thick IBAD YSZ film, and an 80-μm x 1-mm bridge onto a 0.20-μm-thick RABiTS film. The transport measurements were conducted in a four-point strip configuration in a cryogenically cooled 15-T superconducting magnet. Contact resistance was negligible, on the order of a few tens of micro-ohms. To further minimize any joule heating, current is pulsed with a duration of 30 ms, with voltage measurement sampled at the center of the pulse. To eliminate any thermal voltage offset, current is pulsed in opposite directions and the voltages averaged. Since the previous study showed the hysteresis occurring at very low fields, any fields trapped within the cryostat magnet, typically around 50 G, could limit the resolution at fields of interest. Therefore, instead of relying on the set field, a Hall probe was used to independently measure the magnetic field at the sample.

## 3. Discussion of Results

First we consider the films patterned into 100-µm parallel strips. The goal was to attempt to isolate one or a few GB's along the current flow path – the likelihood of which would be controlled by the grain *size* of the film. Since IBAD grains are known to be much smaller than those of RABiTS, across a 100-µm width one expects between ten and 100 IBAD grains, and only a few RABiTS grains (possibly one). Thus we fully expect the effect to be more apparent in the RABiTS sample. This is indeed the result, shown in Figure 1. For the IBAD case (1a), the maximum separation between $J_c$ (H) for increasing and decreasing H (at around 300 G) represents a difference of about 8%, while for the RABiTS case (1b), the separation is closer to 28% (at about H = 400 G). In addition, one sees a $J_c$ "peak" in both cases; however, the field $\underline{H_{peak}}$ where this occurs is much lower in the IBAD sample, at about 80 G, than in the RABiTS sample in which $J_c$ peaks at around 200 G. This strongly implies that a higher-angle grain boundary – or an effective combination of GB's – was isolated after the patterning, drawing out the granular hysteretic effect. This seems consistent with the reduction of the overall $J_c$ by more than an order of magnitude after patterning, implying an isolated GB of significant misorientation. It is well known that a tenfold reduction in $J_c$ can occur at these low fields for a misorientations less than $10°$[2].

It is interesting that the IBAD sample showed the hysteretic effect, since there are nominally as many as 100 grains across the width, based on the substrate morphology. Another IBAD film was patterned to a much wider bridge of 200 µm and the effect was similar, as shown in Figure 2. If anything, the effect seems even enhanced, with a wider separation between increasing- and decreasing- field branches, although the $J_c$ peak is similar. However, this may be consistent with the fact that there is simply very little difference between a width reduction of 98% (from 4000 µm to 80 µm) and 95% (from 4000 µm to 200 µm). For the full-width sample, the curves for $J_c$ (H) for increasing field (right-pointing triangles) and decreasing field (left-pointing arrows) overlap very well. It should be noted that there is also an apparent uniform decrease in $J_c$(H) after patterning: the full-width data has been plotted after dividing by a factor of 1.37. Since this scaled data merges with after-patterned data at high fields (where $J_c$ is grain-limited), we speculate that there had been a finite range of degradation at the outer edges of the bridge, causing the effective width to be actually narrower and the corresponding $J_c$ to be slightly under-calculated. Nevertheless there is a discernible peak for the decreasing field; however this is quite broad compared to single-GB results [11], which would be consistent if flux trapping were occurring over a *range* of fields *H*, over which many peaks $H_{peak}$ would therefore be averaged. In addition, the data eventually overlap for higher fields (above 2000 G). This is consistent with the well-established observation [23, 26] that at higher fields *intra*granular effects are more dominant than intergranular effects: At higher fields depinning of Abrikosov vortices within the grains limits the overall critical current.

However, by virtue of increased flux pinning, flux trapping is stronger at lower temperatures; consequently one expects the hysteresis to be enhanced as temperature is lowered. This is indeed what is observed for IBAD, Fig. 3a, for temperatures 77 K, 60 K, and 40 K. For the same reason, the field $H_{peak}$ at which the flux-trapping peak occurs is shifted towards *higher* fields at lower temperatures. An even better example of this is shown for a film on RABiTS, Fig. 3b, narrowed down to 80 μm. Consequently the flux-trapping peaks are much sharper in this case than with the case of IBAD-YSZ above. Compared to single-GB results, however, these peaks are still relatively broad, implying effects of averaging over multiple grain boundaries, or variations in field within the grain boundary. Also, the field $H_{peak}$ at which the peak occurs increases with decreasing temperature. Again, this is consistent with the expected improvement of flux trapping in the grains and enhancement of the circulating intragranular $J_c$ at lower temperatures. Another feature in these plots is the fact that the $J_c(0)$ value is the same as the value of $J_c$ at the peak ($H = H_{peak}$) for the decreasing field branch, suggesting that the field $H_{peak}$ has nearly perfectly cancelled the GB field, so that the $J_c(H_{peak}) \cong J_c(0)$, where $J_c(0)$ is influenced only by the current self field.

Further consistency with the flux-trapping model is seen by re-plotting $J_c(H)$ in log-log, shown in Figure 4. In other flux pinning studies [27] it has been established that at intermediate fields, on the order of 0.1 to 1 T at 77 K, there is an observed power-law regime where $J_c(H) \sim H^{-\alpha}$; the value of the exponent α has been associated with various models for flux pinning. For our particular case, the value is around 0.6, close to the value of 5/8 which is predicted by Ovchinnikov and Ivlev [28] for large, dilute pinning defects, such as isotropic precipitates. This has been found consistent with as-made *ex situ* films[24, 29, 30], and is very likely the mechanism by which vortices are trapped in the intragranular phase. In Figs. 4 and 5, for the narrowed case the exponent is the same for both increasing and decreasing field; however the range of fields is narrower for *increasing* field. This would be consistent with the presumption that, with *increasing* applied field, the field in the GB is a combination of both applied field and the trapped field in the grains, which would suppress $J_c$ up to higher fields. It is also important that the two curves merge at higher fields, where one expects intragranular current to be limiting. Incomplete penetration of field within the grains is also possible for increasing field, but this would be likely only at lower applied fields ≤ 1 kG [31]. The effect is most pronounced for the 80-μm RABiTS case of Figure 3(b), shown in Fig. 4(c). It is also noteworthy that the "cutoff" fields where the power-law regime starts, are very similar for the two species, suggesting that the relative effects of grain-boundary limitations at low field, and flux pinning limitation at high field are similar for the two types of substrates. This may originate by virtue of their similar film processing via the *ex situ* $BaF_2$ method, although the slightly lower α value for RABiTS suggests a somewhat improved pinning mechanism than that on IBAD-YSZ.

**4. Conclusions**

This work has shown that narrowing coated-conductor film on RABiTS or on IBAD-YSZ "pinches off" the percolative current enough to manifest the hysteresis of transport $J_c$ due to granularity. The result is that $J_c(H)$ is generally *higher* when the field is decreased, but only down to the "peak" field $H_{peak}$, where the approximate self-field $J_c(0)$ has been effectively shifted due to field focusing at the grain boundaries. As long as coated conductors are *not* single crystals, to properly ascertain $J_c(H)$ at fields well below 2 kG the history by which magnetic field *H* is applied should be taken into account. Furthermore, it is possible that narrow strips isolate GB's of high misorientation and degrade the sample $J_c$. Another practical result is that smaller-grain coated conductors are clearly favorable for patterning into narrow conduits. Also, since the hysteretic effect is governed by the flux trapping capability within the grains, it would be interesting to investigate if this effect is greater in films where pinning is greatly enhanced by pinning-effective nanostructures [32].

**Acknowledgments**

Work at ORNL sponsored by the U.S. Department of Energy - Office of Electricity Delivery and Energy Reliability and by the Office of Science, Division of Materials Sciences and Engineering. A. A. Gapud supported by the University of South Alabama College of Arts and Sciences and the University of South Alabama Research Council.

**Figure Captions:**

**Figure 1.** $J_c$(H) for increasing and decreasing field H, after patterning into 100-μm parallel strips: (a) Sample on IBAD-YSZ and (b) Sample on RABiTS. Both insets show onset of reversible $J_c$ (H) at higher fields as indicated by arrow.

**Figure 2.** $J_c$(H) for increasing and decreasing field H, before (open symbols) and after (solid symbols) patterning an IBAD-YSZ film into a 200-μm-wide bridge via laser scribing. No hysteresis is discernible for the full-width case, in contrast to the narrowed case. The overall $J_c$ is slightly reduced after patterning, most likely due to additional damage at bridge edges, thus effectively narrowing the bridge further.

**Figure 3.** Narrowed films' $J_c$(H) for increasing and decreasing field H, at different temperatures: (a) 200-μm-patterned film on IBAD-YSZ (same sample as in Figure 2); (b) film on RABiTS (same sample as in Figure 1b), further narrowed down to 80 μm.

**Figure 4.** Power-law field dependence of critical current density, $J_c \sim H^\alpha$ for film on RABiTS: (a) at full width, 77 K; (b) after patterning and narrowing down to 200-μm, for different temperatures; and then (c) to 80-μm, same three temperatures. In both full-width and 200-μm-narrowed, $\alpha = 0.6$; arrows show fields for onset of power-law behavior for decreasing field. Note that these fields are unchanged after further narrowing to 80 μm. However, $\alpha$ decreases to 0.5; see text.

**Figure 5.** Power-law field dependence of critical current density, $J_c \sim H^\alpha$ for film on IBAD-YSZ: (a) at full width, 77 K; (b) after patterning and narrowing down to 200-μm, for different temperatures. In both cases, $\alpha = 0.6$; arrows show fields for onset of power-law behavior for decreasing field. Note that these fields are very similar to those in the RABiTS case.

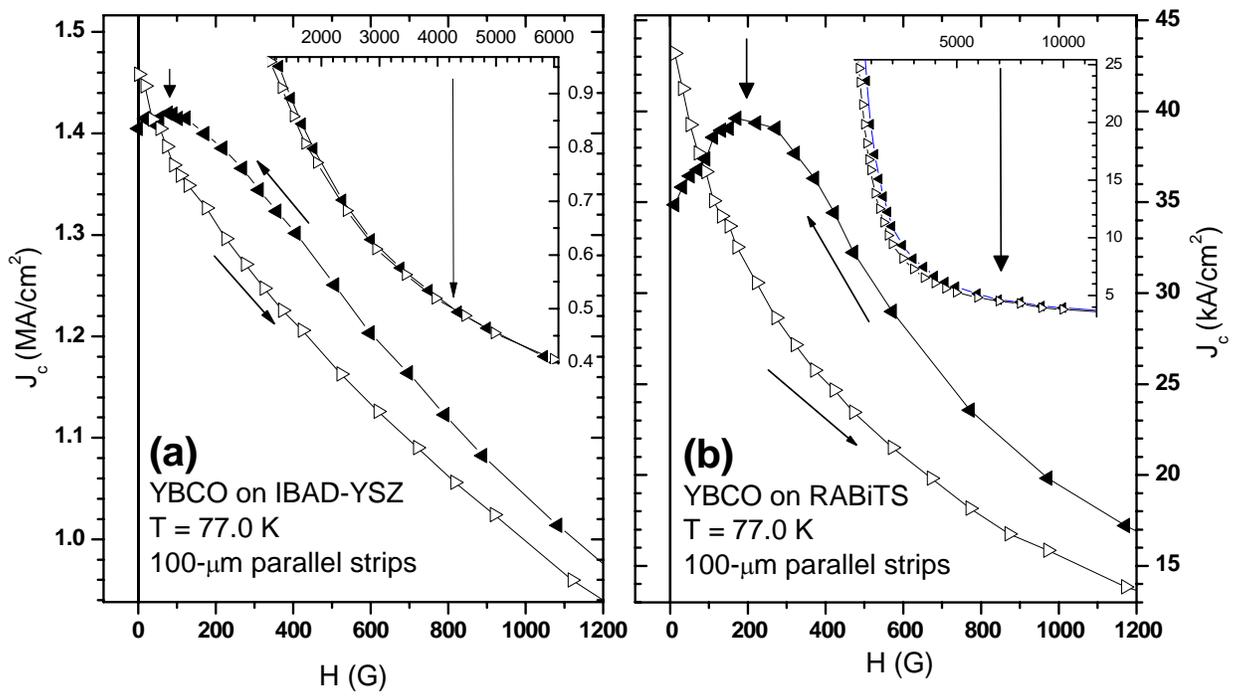

Fig 1 Gapud *et al.*

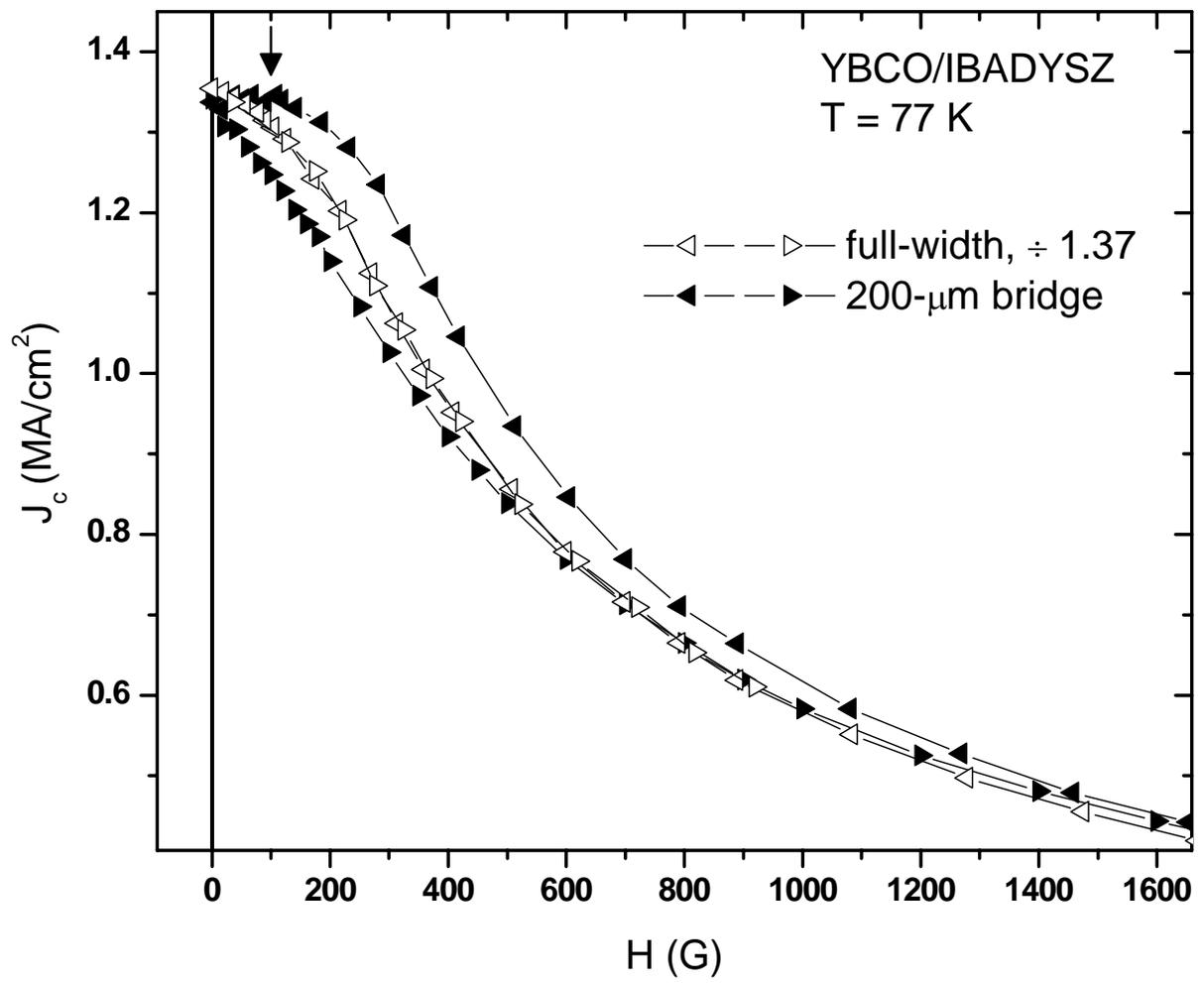

Fig 2  Gapud *et al.*

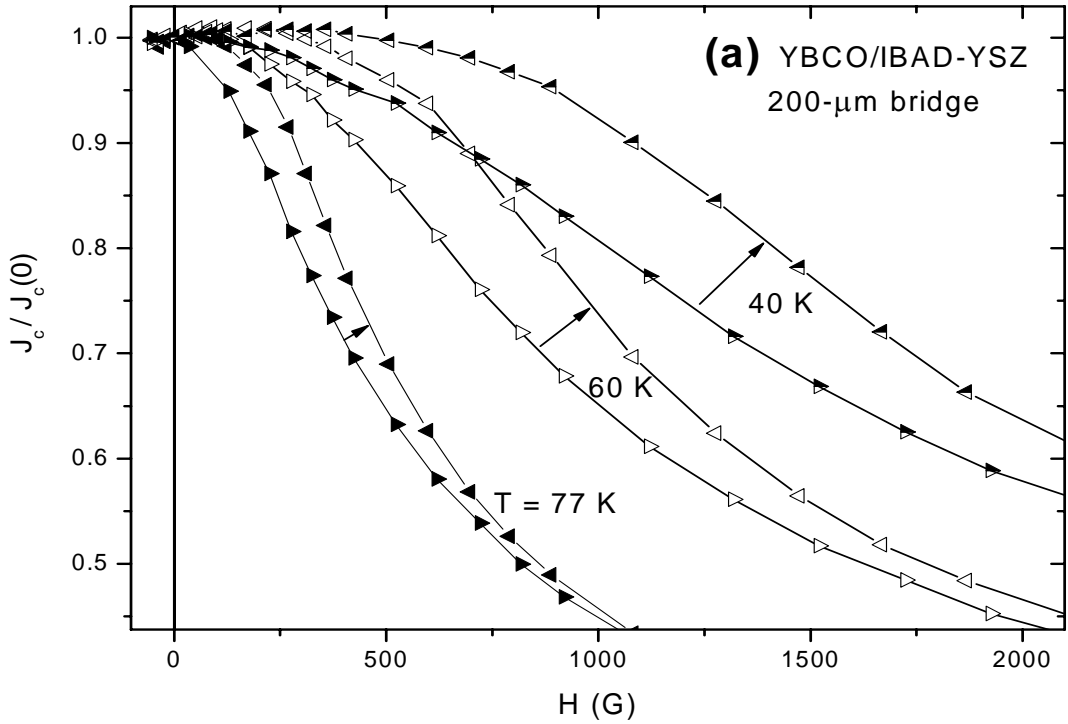
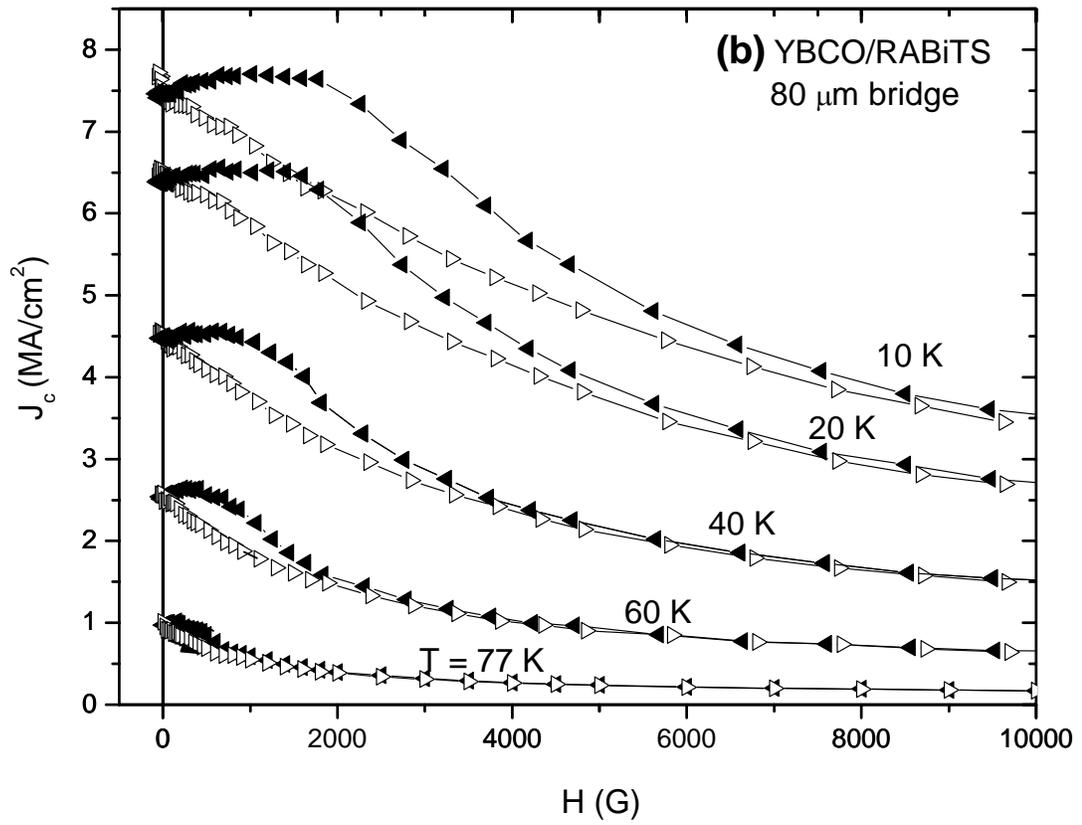

Fig 3    Gapud *et al.*

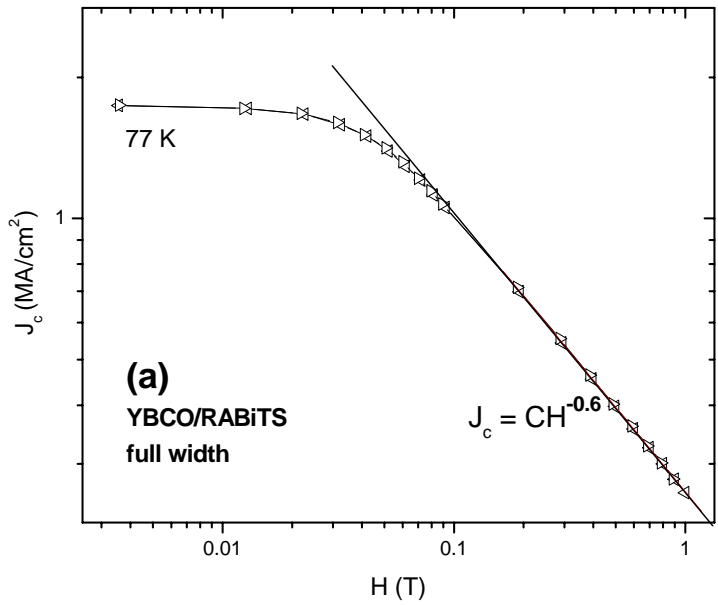
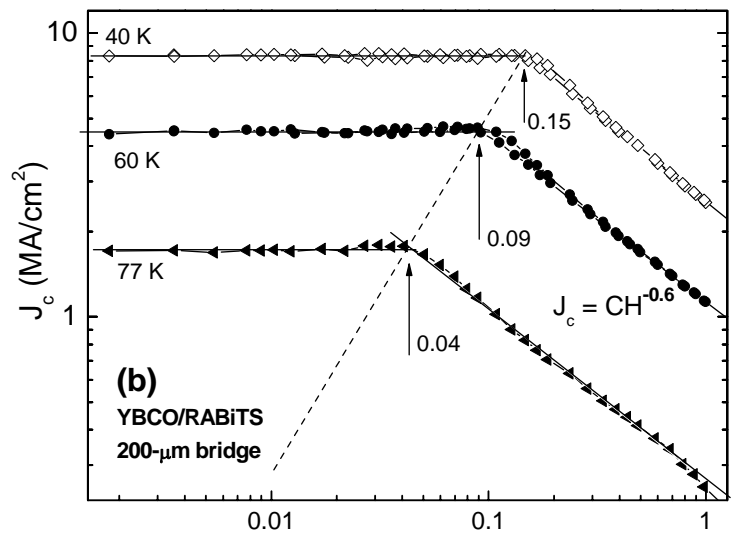
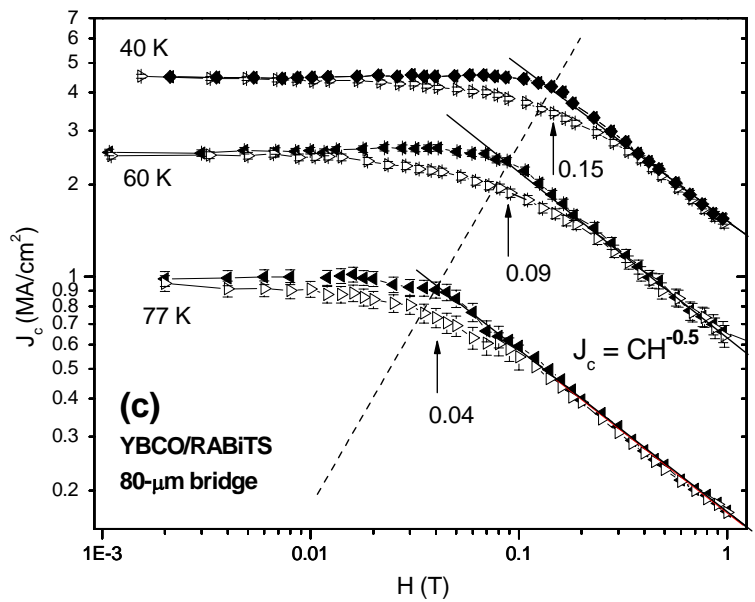

Fig 4   Gapud *et al.*

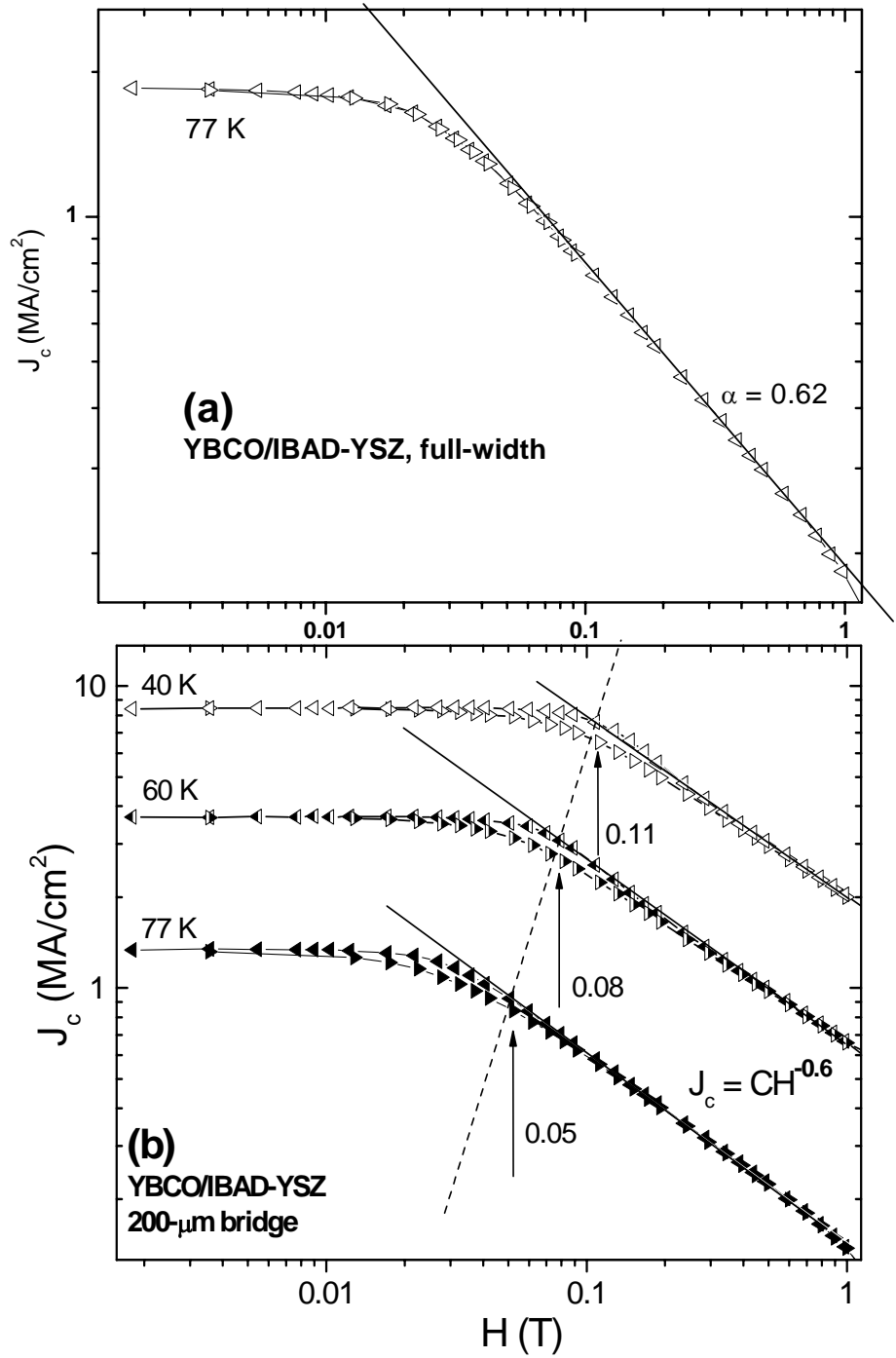

Fig 5   Gapud *et al.*